# "Short-term time prediction" of large EQs by the use of "Large Scale Piezoelectricity" generated by the focal areas loaded with excess stress load.

Thanassoulas, P.C., B.Sc in Physics, M.Sc – Ph.D in Applied Geophysics.

Retired from the Institute for Geology and Mineral Exploration (IGME)
Geophysical Department, Athens, Greece.
e-mail: thandin@otenet.gr – website: www.earthquakeprediction.gr

**Abstract.**

In this work, it is demonstrated that the Earth's preseismic electric field, which is registered by a pair of electrodes in contact to the ground surface at certain distance from the epicentral area, corresponds to the gradient of the total field generated in the focal area as a function of time and distance from it. The original form of the generated preseismic field follows closely the theoretical piezoelectric potential form. The later is obtained after integration in time of the original registered potential grad data values.

Consequently, the time of occurrence of the imminent earthquake (collapsing of the rock formation) is estimated by the classical laws of rock fracturing based on theoretical Rock Mechanics.

The methodology has been tested on the grad potential data, registered in Greece, before four large EQs which took place in the regional area as follows: Izmit, Turkey EQ (M = 7.8R, 17$^{th}$ August, 1999), Milos, Greece EQ (M = 5.6R, 21$^{st}$ May, 2002), Kythira Greece EQ (M = 6.9R, 8$^{th}$ January, 2006) and Methoni, Greece EQ (M = 6.7R, 14$^{th}$ February, 2008).

The obtained results prove the validity of the postulated methodology.

## 1. Introduction.

Concerning the preseismic electrical signals, it is well known, that various physical mechanisms do exist in account of their generation (Thanassoulas, 2007 and therein references) and that the generated preseismic signal amplitude depends upon its wavelength (period / frequency). Signals such as train pulses (SES), oscillating signals mainly of a day's period, very long period (of some days) signals have been observed frequently some time before the occurrence of a large earthquake.

These types of preseismic signals have been used either for determination of the time of occurrence of an imminent large EQ or for the determination of its location (Thanassoulas, 2007).

The referred signals are received by appropriate electrode arrays, which are used for their registration. The questions which arise immediately are:

  a. Which physical quantity is measured on the ground surface by using any electrical array (dipole)?

  b. How the preseismic electric field can be related to the time of occurrence of the imminent large EQ?

These questions will be answered in detail, in a quite different way from what has been presented today, following a new approach as follows.

## 2. The theoretical model.

Let us assume that a seismogenic area has been activated and generates, by any acceptable and valid physical mechanism, a potential **(P)** at its focal area. The potential **($P_1$)**, exactly above its focal area, at the epicenter, will exhibit slightly lower amplitude than **(P)**, due to the fact that the epicenter is located at some distance **(x),** far from the focal area and moreover the potential **(Pr),** at any distance **(r)** from its origin, will be a function **(f)** depending on distance **(r)** and the time of observation **(t)**.

$$Pr = f(r,t) \qquad (1)$$

Adopting the simplest current generating model, the **point current source** (Thanassoulas, 1991), as the generating mechanism, the potential **(Pr),** as a function of distance from the epicentral area, is shown in the following figure **(1)**.

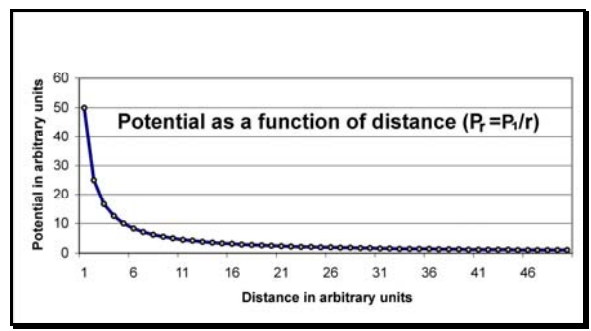

Fig.1. Schematic presentation of the potential **(Pr)** at a location, as a function of its distance **(r)** from the epicenter area of a pending, strong earthquake, is shown.

The registration of the Earth's potential is made through the use of a pair of electrodes, forming a dipole of length **(dx),** in contact to the ground. Therefore, what is actually measured is the potential field quantity:

**dP = ∂P/dxdt**  (2)

assuming that:

**dx ► dr**  (3)

then equation **(2)** can be written as:

**dP = ∂P/drdt**  (4)

that is the time / distance gradient of the potential field which is generated in the focal area.

Starting from equation **(4)** it is possible by integration to calculate either: a) <u>the variation of the potential as a function of time</u>, by keeping constant the **(r)** value, or b) <u>the variation of the potential as a function of distance **(r)**</u> by keeping constant the time.

At any time **(t)** and distance **(r)**, figure **(2)** indicates the quantity, which is measured by the used dipole.

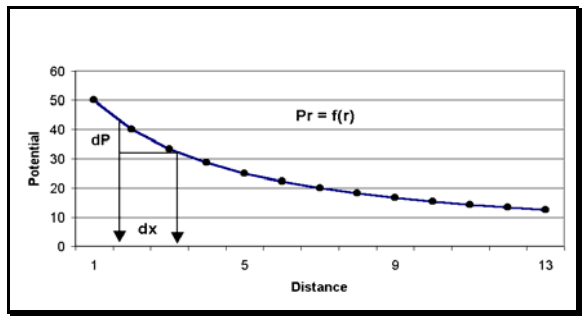

Fig. 2. Graphic presentation of the quantity **(dP or ΔP),** measured at any time **(t)** and distance **(x)**, by a pair of electrodes, which are inserted in the ground at distance **(dx)** between each other.

The true form of the generated potential, as a time-function, is obtained after integrating the equation **(4)** in time. The later is achieved by successive summation of the, sequential in time, measured **ΔP** values, at **dt (Δt)** time intervals (equation **5**).

$$P(t) = \sum_{0}^{t} \Delta P_{t_i}$$  (5)

A pictorial presentation of the operation of equation (**5**) is presented in the following figure (**2a**).

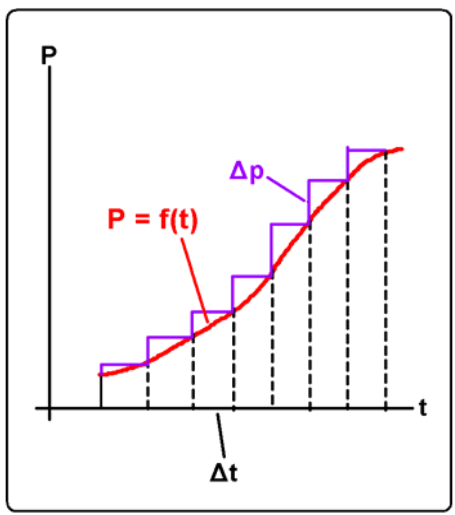

Fig. 2a. Recovery of $P = f(t)$ after integration in time, from measured **ΔP** values at **Δt** intervals.

The integration constant of equation (**5**) is set arbitrarily as **0,** since we are interested in long term, in time, potential differences only.

## 3. Model from Rock Mechanics

At this point, let us recall, the stress-strain relation of a stress charged, deforming solid and the corresponding potential-time curve, which is generated when the stressed material exhibits piezoelectric-like properties. This is presented in the following figure **(3)**.

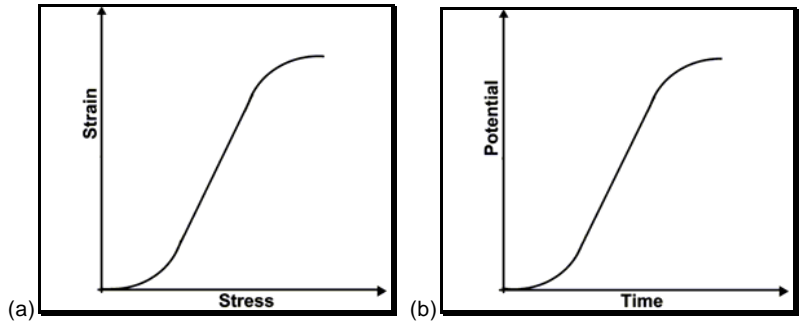

Fig. 3. Typical stress – strain relation of solid material that exhibits piezoelectric-like properties and its corresponding, potential – time relation. It is assumed that positive charges are generated in compression mode.

Although figure (**3**) represents the generalized case of the piezoelectric mechanism, the following sub-cases can be activated which depend on parameters such as: applied compression or decompression in relation to the generation of positive or negative electrical charges. The later are presented in the following figures (**3a, 3b, 3c**).

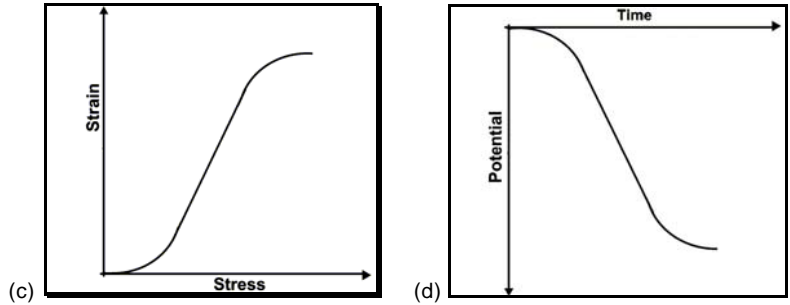

Fig. 3a. The same as in fig. (3) but negative charges are generated instead in compression mode.

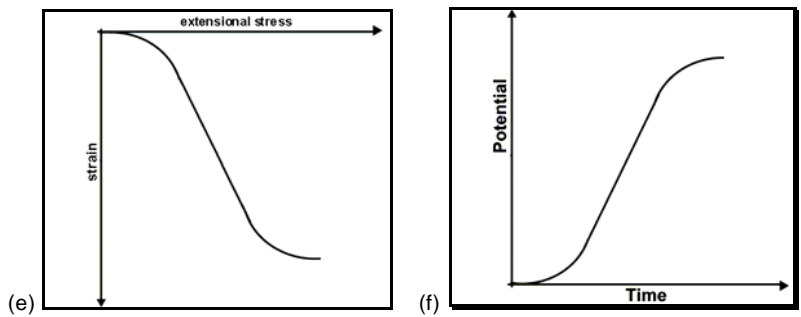

Fig. 3b. Extension stress is applied and positive charges are generated in extension mode.

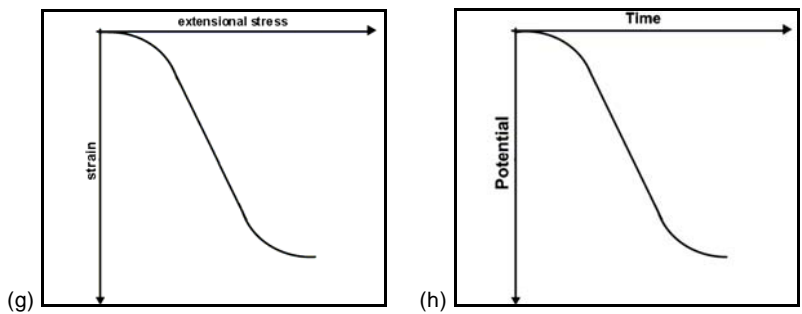

Fig. 3c. Extension stress is applied and negative charges are generated in extension mode.

Furthermore, the polarity of the observed field depends on the location, in a Cartesian coordinating system, of the monitoring site in relation to the location of the charges generating mechanism.

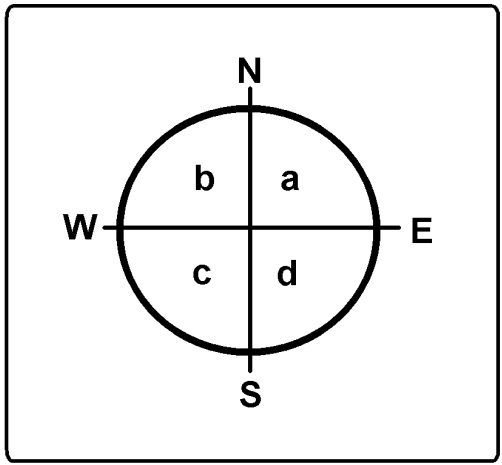

Fig. 4. Trigonometric circle indicating the four quadrants which relate the observed polarity of the registered electric field of a monitoring site to the location of the generating physical mechanism (monitoring site – epicentral area relative location).

Assuming a compression stress mode and a positive charge generating mechanism then, the total field form obtained by integration of the grad electric field conforms to the following:

1. **EW** and **NS** are positive components (**quadrant a**): fig. 3 is valid for both **NS** and **EW** calculated total fields.

2. **EW** is negative and **NS** is positive component (**quadrant b**): fig. **3** is valid for **NS** and fig. **3a** is valid for **EW** calculated total fields.

3. **EW** and **NS** are negative components (**quadrant c**): fig. **3c** is valid for both **NS** and **EW** calculated total fields.

4. **EW** is positive and **NS** is negative component (**quadrant d**): fig. **3** is valid for **EW** and fig. **3a** is valid for **NS** calculated total fields.

The figures **(b, d, f, h)** on the right side of figure **(3, 3a, 3b, 3c)** represent the total potential which is generated by the well-known in physics piezoelectric mechanism.
By keeping the (**x**) distance as constant, integration in time will reveal the true form of the generated potential as a function of time.
The registered on ground surface, potential gradient which is generated by any activated, physical mechanism in the focal area, after its integration in time, will present its original, potential form. Consequently, if the resulted potential, after this operation (integration) presents the form of the right part of figures **(3, 3a, 3b, 3c),** then it is justified to accept that the main, generating mechanism is the piezoelectric one, activated in the crust, mainly due to the presence of the quartzite and stress increase.
In practice, the polarity of the generated electrical field, in the focal area, depends on parameters still unknown, which must be studied in detail in the future. Nevertheless, the knowledge of the polarity of the generated field is not important, as far as it concerns the present methodology.
The issue of the piezoelectric mechanism has been objected strongly by many researchers.
**The main objection is that, quartzite crystals are randomly oriented, in space in the lithosphere and therefore, the potentials which are generated by single crystals are cancelled out by others, of opposite polarity.** This leads to a neutral, electrical property of the lithosphere. This point of view is strongly debated by the fact that, during crystallization, various factors affect the orientation of the main axis of the quartzite crystals.
Probably, the main factor is the stress, present in the media of crystallization. Another one is gravity and weight load charge, due to overlaying, geological formations. Large tectonic movements, also, change the crystal axis orientation of the quartzite crystals, which are contained in large, geological blocks. Moreover, even if it is accepted that there is a smooth, random orientation of the quartzite crystals, still only the crystals which have their main symmetry axis, aligned, to the stress orientation axis will generate piezoelectric potential. In this case, it is very difficult to adopt the idea that there is a "perfect" match, of the opposite polarity of the activated crystals, which cancels out the generated potential. This is especially more difficult, if the fact that the crust is not a perfect, homogeneous, electrical medium, is taken into account. Furthermore, the piezoelectric properties of the quartzite are so large, compared to other materials, that even a small amount of "perfect, main symmetry axis orientation" mismatch will produce potentials, observable, on ground surface at large distances. The reason which justifies the use of quartzite in various applications by the industry, from crystal oscillators in radio transmitters to small lighters, is exactly its large piezoelectric property, compared to any other.
Another objection, arising, against piezoelectricity is that, the generated potentials fade out rapidly, because of Earth's high conductivity. The work of Blohm (1977), who suggests a highly resistant crust (100000 Ohm.m), at the main, seismogenic zone, when we refer to the conductivity of the Earth, must be recalled. The presence of the piezoelectric, generating mechanisms, which are activated in the focal area, in a highly resistant medium, facilitates the generated potentials to be sustained furthermore and to propagate at long distances with no appreciable attenuation. Therefore, the argument of the conductive Earth does not apply in the seismogenic zone of the crust.
Apart from these, "for and against" arguments, piezoelectricity is the only, known mechanism, which justifies the generation of:

a) Long period oscillating signals (T = 1, 14 days), due to the fact that it responds to the (induced by the tidal waves) strain deformation of the, subjected, in stress, material, as long as the stress lasts,

b) VLP signals which are related to the first time derivative of the generated piezoelectric potential.

c) SES due to the piezostimulated currents generated in the non-linear parts of the piezoelectric potential as a function of time, specifically during the onset of the piezoelectricity mechanism and just before the occurrence of the large EQ by rock fracturing / collapsing.

The already presented theoretical analysis can be applied on the registered electric field measured on ground surface at any monitoring site.

If the focal area of a future large EQ has entered its final phase of drastic rock deformation then it must exhibits piezoelectric-like behavior.
Consequently, it could be possible to monitor the form of the generated potential and estimate when the focal area has reached its collapsing stress – strain level, by comparing it to the theoretical one, thus suggesting the most probable time of occurrence of the pending large EQ in a "short-term" mode.

This method will be applied on real electric signals recorded in Greece before the occurrence of large seismic events.

## 4. Examples from real EQs.

The examples which will be presented in the following text have been calculated from real actual grad measurements of the Earth's electric field at VOL, ATH, and PYR monitoring sites (Thanassoulas, 2007, www.earthquakeprediction.gr).

### 4.1 Izmit, Turkey EQ (M = 7.8R, 17th August, 1999).

The preseismic, electrical signals were recorded at a distance of 650Km from Izmit, at **VOL**, Greece, with a dipole of **120m**, oriented, **NS**. The relative location of Izmit – VOL corresponds to quadrant (a).
The registered, raw data (grad) are presented in figure **(5)**.

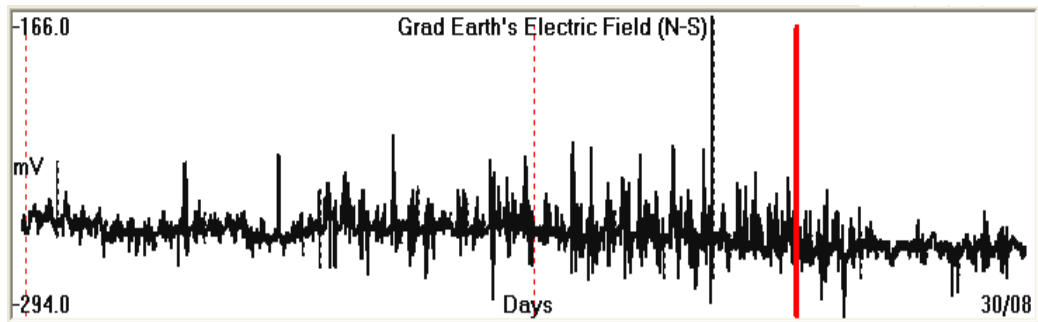

Fig. 5. The Earth's electric field (grad), as it was recorded at **VOL**, Greece, for a two months period (1st July – 30th August 1999). The red bar indicates the time of occurrence of Izmit, Turkey EQ (M = 7.8, 17th August, 1999).

The recorded signal indicates some "noise" increase for some days before the occurrence of the EQ. Regardless the nature of this "noise", the raw signal must be detrended, so that the superposed DC component is eliminated.
Finally, integration is performed along the detrended data and the original potential, which was generated by the physical mechanism, triggered in the focal area, has the form of the figure **(6)**.

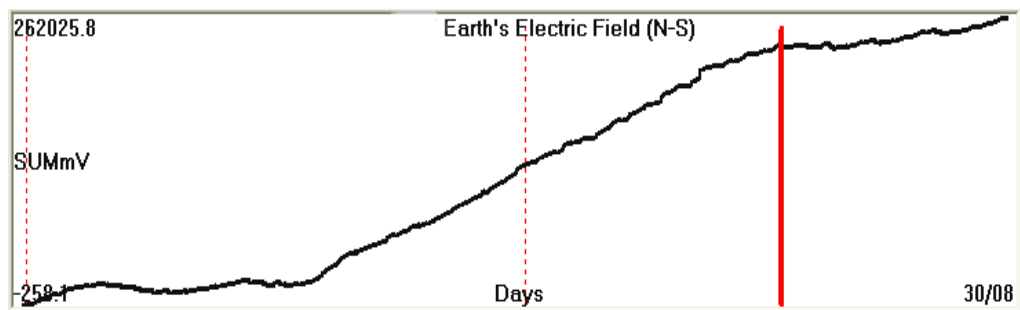

Fig. 6. Form of the potential, generated, by the physical mechanism, triggered, in the focal area of Izmit, Turkey EQ (red bar, M = 7.8, 17th August, 1999).

Next, figure **(7)** compares the potential results, which are obtained, from the processing of the Izmit, registered, grad data to the theoretical ones, which are indicated, by the adopted, piezoelectric mechanism.

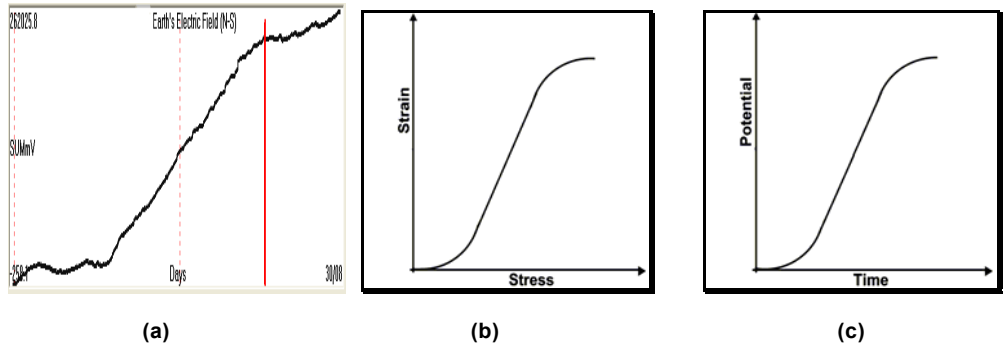

(a)                  (b)                  (c)

Fig. 7. Potential **(a)** generated by the EQ focal area, compared to the theoretical **(c)** piezoelectric model and the triggering **(b)** stress, inducing mechanism.

It is worth to notice that the earthquake occurred, as it was predicted by the stress-strain curve following the rock-mechanics fracturing laws.

### 4.2 Milos, Greece EQ (M = 5.6R, 21$^{st}$ May, 2002).

The raw data which were recorded at a distance of 314 Km from Milos epicentral area, from **VOL**, for a period of five months (31$^{st}$ December 2001 – 1$^{st}$ June 2002), by a dipole of 120m oriented **NS,** are shown in the following figure (**8**). The relative location of Milos - **VOL** corresponds to quadrant (**c, d**).

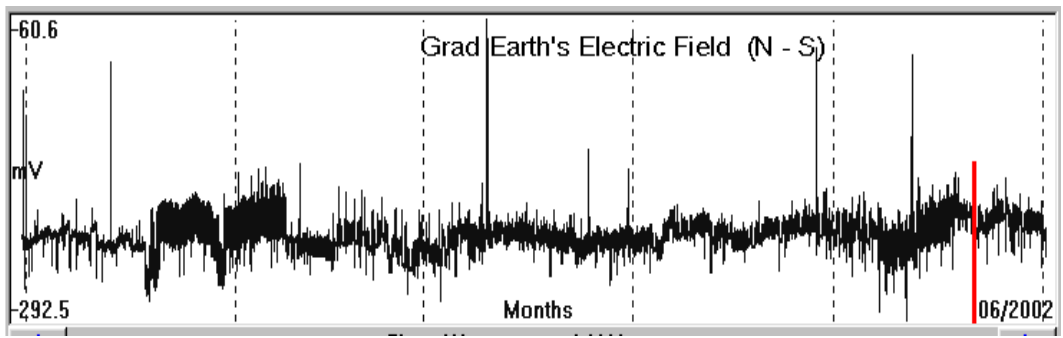

Fig. 8. The Earth's electric field (grad) as it was recorded in Volos **(VOL)**, Greece for a five months period (31$^{st}$ December, 2001 – 1$^{st}$ June, 2002). The red bar indicates the time of occurrence of Milos, Greece EQ (M =5.6, 21$^{st}$ May, 2002).

The recorded signal indicates some "noise" increase for some days before the occurrence of the EQ and a similar one 3.5 months before it. The same, detrending operation was applied in these data, too, so that the superimposed DC component is eliminated.

Finally, integration is performed along the detrended data and the generated, original potential, by the triggered, physical mechanism, in the focal area, has the form of the figure **(9)**.

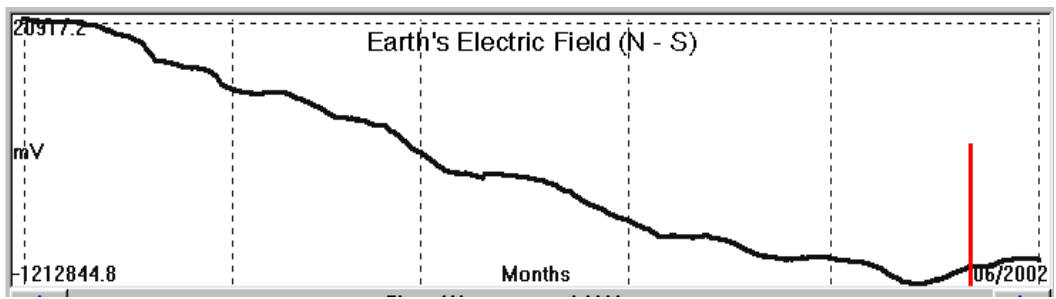

Fig. 9. Form of the potential, generated, by the triggered, physical mechanism, in the focal area of Milos, Greece EQ (red bar, M =5.6, 21$^{st}$ May, 2002).

Next, figure **(10)** compares the potential results, which are obtained from the processing of Milos, Greece EQ, registered grad data, to the theoretical ones, indicated by the adopted, piezoelectric mechanism.

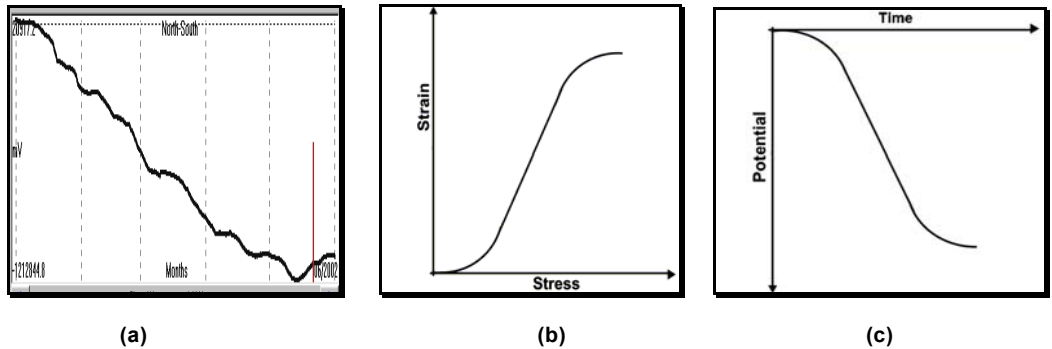

Fig. 10. Potential **(a)** generated by the EQ focal area, compared to the theoretical **(c)** piezoelectric model and the compression mode **(b)** triggering stress, inducing mechanism.

The decrease of potential, observed, in **(a)** is due to the fact that, either negative polarity potential is generated in the focal area or this is the effect of "quadrant c, d" VOL – Milos relative location. The typical form of figure **(c)** is based on the assumption that positive polarity potential is generated.

### 4.3  Kythira, Greece EQ (M = 6.9R, 8$^{th}$ January, 2006).

The raw grad data which were recorded, at a distance of 239Km from Kythira with dipoles of 160m length oriented **NS** and **EW**, by **PYR** monitoring site, after normalization, are presented in figure **(11)**. The relative PYR – Kythira location suggests quadrant (**d**).

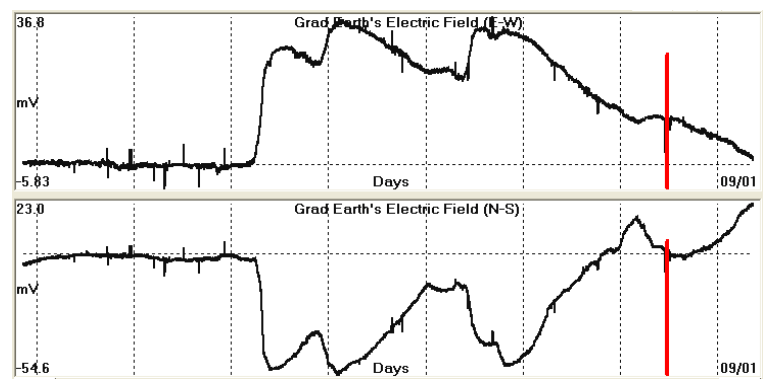

Fig. 11. The Earth's electric field (grad) as it was recorded in **PYR**, Greece for a seven days period (2$^{nd}$ – 8$^{th}$, January, 2006). The red bar indicates the time of occurrence of Kythira, Greece EQ (M = 6.9R, 8$^{th}$ January, 2006).

The application of the methodology on the data of fig. (**11**), has as a result the potential which is presented in the following figure (**12**).

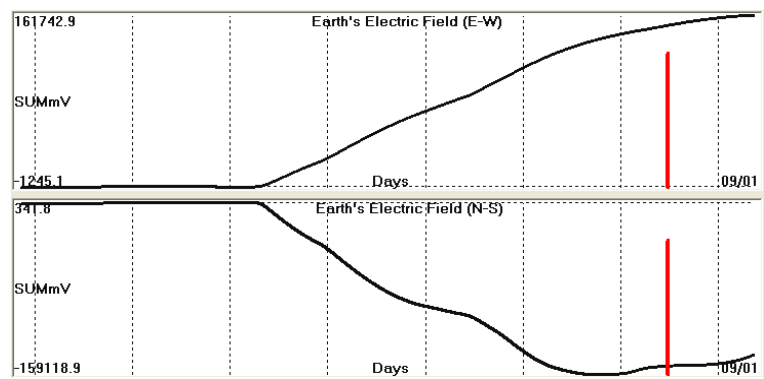

Fig. 12. Form of potential (**NS** and **EW** components), generated, by the triggered, physical mechanism, in the focal area of Kythira, Greece  EQ (M = 6.9R, 8$^{th}$ January, 2006). The red bar indicates the time of occurrence of Kythira,  EQ.

Next, the theoretical potential – time curves are compared to the graph of figure (**12**). This comparison is presented in the following figures (**13, 14**).

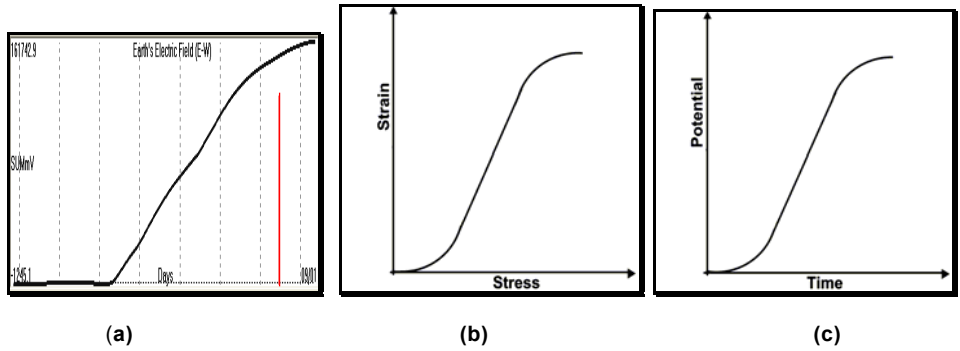

(**a**)          (**b**)          (**c**)

Fig. 13. Potential (**a, EW** component) generated by the EQ focal area**,** compared, to the theoretical **(c)** piezoelectric model and the compression **(b)** mode triggering stress, inducing mechanism**.**

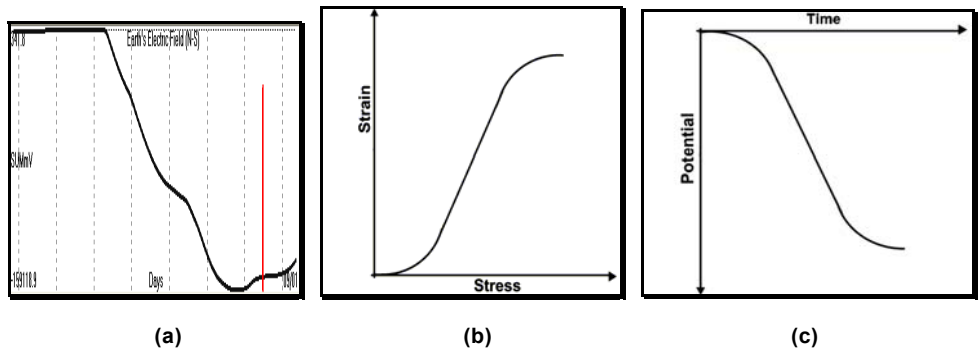

(**a**)          (**b**)          (**c**)

Fig. 14. Potential (**a, NS** component) generated by the EQ focal area**,** compared to the theoretical **(c)** piezoelectric model and the compression **(b)** mode triggering, stress inducing, mechanism**.**

### 4.4 Methoni, Greece EQ (M = 6.7R, 14th February, 2008).

The raw grad data which were recorded at a distance of 257Km from Methoni, with dipoles of 21m length oriented **NE** and **NW**, by **ATH** monitoring site for a period of about 3 months (14st December, 2007 – 21st February, 2008) after normalization are presented in figure **(15)**. The relative location of **ATH** – Methoni suggests quadrant (c).

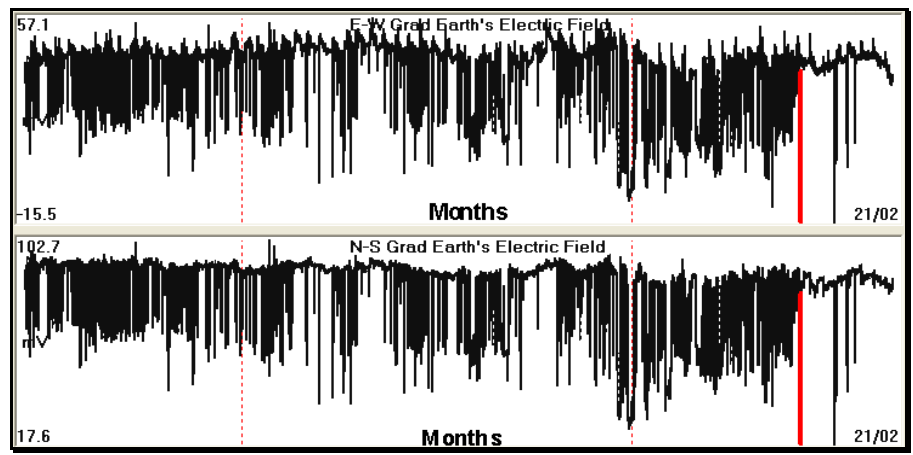

Fig. 15. The Earth's electric field (grad) as it was recorded in **ATH**, Greece for a 3 months period (14st December, 2007 – 21st February, 2008). The red bar indicates the time of occurrence of Methoni, Greece EQ (M = 6.7R, 14th February, 2008).

The application of the methodology on the data of figure (**15**) has as a result the potential which is presented in the following figure (**16**).

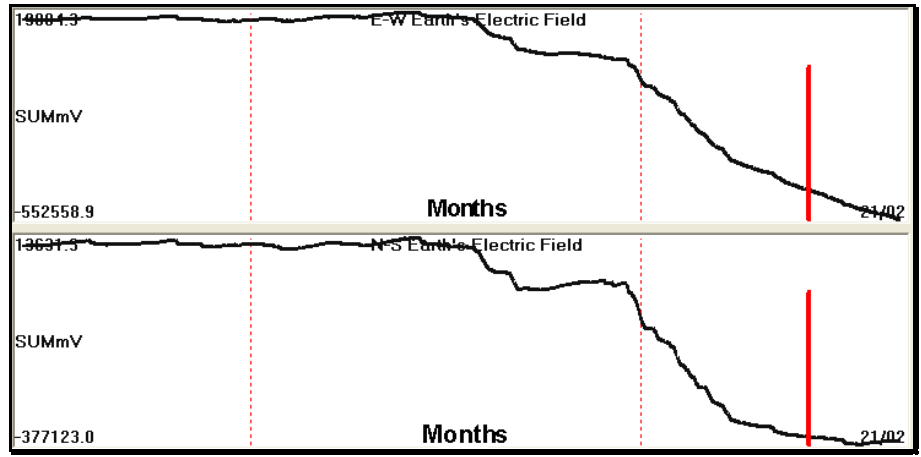

Fig. 16. Form of potential (**NS** and **EW** components), generated, by the triggered, physical mechanism, in the focal area of Methoni, Greece EQ (M = 6.7R, 14[th] February, 2008). The red bar indicates the time of occurrence of Methoni EQ.

Next, the theoretical potential – time curves are compared to the graph of figure (**16**). This comparison is presented in the following figures (**17, 18**).

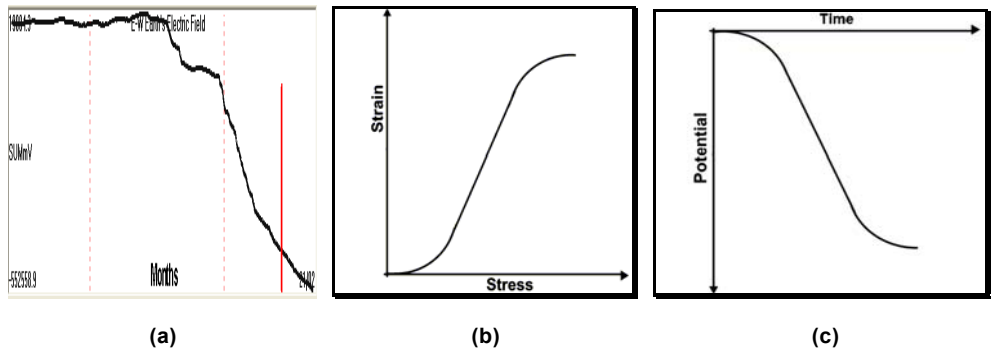

(a)   (b)   (c)

Fig. 17. Potential (**a, EW** component) generated by the EQ focal area, compared to the theoretical **(c)** piezoelectric model and the compression **(b)** mode triggering stress, inducing mechanism.

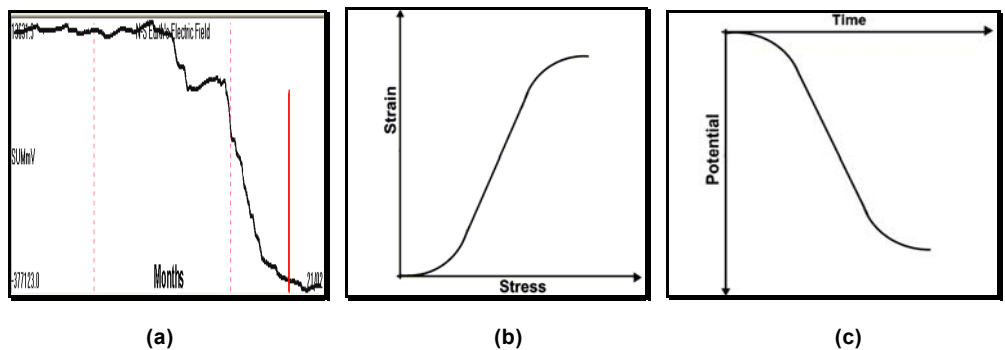

(a)   (b)   (c)

Fig. 18. Potential (**a, NS** component) generated by the EQ focal area, compared to the theoretical **(c)** piezoelectric model and the compression **(b)** mode triggering stress, inducing mechanism.

Finally, a comparison is made between the registered Earth's electric field polarity at each monitoring station and the suggested electric charge polarity assumed for each trigonometric quadrant (a, b, c, d) of figure (4). The later has been utilized by taking into account the relative location of each monitoring site and the corresponding EQ. Thus, for each studied EQ the following charge polarity is implied:

    a.   Izmit EQ        : positive electric charges were generated at the focal area. Quadrant (**a**).

    b.   Milos EQ       : positive electric charges were generated at the focal area. Quadrant (**c** and **d**).

    c.   Kythira EQ     : positive electric charges were generated at the focal area. Quadrant (**d**).

d.  Methoni EQ     : positive electric charges were generated at the focal area. Quadrant (**c**).

It seems that, at least in these studied cases, the very same piezoelectric mechanism was triggered, some time before the occurrence of each earthquake. Moreover, the large EQ of each case did take place as it is suggested by the rock fracturing theory based on rock mechanics. **In other words, the imminent EQ had announced its occurrence short before its actual one**. Therefore, this methodology could be helpful for "short-term" time earthquake prediction. Its application is straight forward; it is just the monitoring of the Earth's electric field and studying its form in terms of the theoretical generated piezoelectric potential one.

**5. Conclusions.**

The following comments can be made in conclusion:

- It was made quite clear that the large scale piezoelectricity is the generating mechanism of the observed preseismic Earth's electric potential in all presented cases.
- By comparing the relative location of each monitoring site to the epicentral location of each large EQ, a positive generated electric charge is observed. Thus, it is postulated that the same driving physical macroscopic mechanism was triggered in all presented cases.
- The imminent EQ takes place when the electric potential reaches its saturation level which, in turn, corresponds to the time just before the rock fracturing. In other words the rock formation has enter, in terms of rock-mechanics, its final non-linear deformation phase.
- The duration of the activation of the piezoelectric mechanism is variable in time. This probably depends upon the stress change rate applied on the rock formation in the focal area.
- However, the saturation potential level was clearly observable in all cases short in advance before the EQ occurrence. This specific capability, to identify the last "non-linear" deformation phase of a rock formation is the used criterion for the "short-term earthquake time prediction".
- The methodology provides a valuable tool for "short-term earthquake prediction" of the time parameter of an imminent large EQ.
- The time of occurrence can be refined further more by the use of other preseismic signals and methodologies (Thanassoulas, 2007).

**6. References.**

URL: www.earthquakeprediction.gr